\documentclass[aps,10pt,twocolumn,superscriptaddress]{revtex4}

\usepackage{amssymb,amsmath}
\usepackage{subfig}
\usepackage{graphicx}
\usepackage{color}
\usepackage{hyperref}
\usepackage{listings}
\usepackage{ragged2e}
\usepackage{mhchem}
\usepackage{tabu}
\usepackage{qcircuit}

\hypersetup{
    bookmarks=true,         
    unicode=false,          
    pdftoolbar=true,        
    pdfmenubar=true,        
    pdffitwindow=false,     
    pdfstartview={FitH},    
    pdfauthor={Armaos, Badounas, Deligiannis, Lianos},     
    colorlinks=true,       
    linkcolor=blue,          
    citecolor=blue,        
}
\graphicspath{{figures/}}

\begin{document}

\definecolor{dkgreen}{rgb}{0,0.6,0}
\definecolor{gray}{rgb}{0.5,0.5,0.5}
\definecolor{mauve}{rgb}{0.58,0,0.82}

\captionsetup{justification=justified,singlelinecheck=false,labelfont=large}

\lstset{frame=tb,
  	language=Matlab,
  	aboveskip=3mm,
  	belowskip=3mm,
  	showstringspaces=false,
  	columns=flexible,
  	basicstyle={\small\ttfamily},
  	numbers=none,
  	numberstyle=\tiny\color{gray},
 	keywordstyle=\color{blue},
	commentstyle=\color{dkgreen},
  	stringstyle=\color{mauve},
  	breaklines=true,
  	breakatwhitespace=true
  	tabsize=3
}

\title{Efficient Parabolic Optimisation Algorithm for adaptive VQE implementations}

\author{V. Armaos}
\email[Corresponding author: ]{bill@pidust.com}
\affiliation{Laboratory of Atmospheric Physics, Department of Physics, University of Patras, Greece}

\author{Dimitrios A. Badounas}
\affiliation{Department of Material Science, University of Patras, Greece}

\author{Paraskevas Deligiannis}
\affiliation{Department of Experimental Physics, CERN, European Organization for Nuclear Research, Geneva, Switzerland}

\author{Konstantinos Lianos}
\affiliation{Department of Computer Engineering \& Informatics, University of Patras, Greece}

\author{Yordan S. Yordanov}
\affiliation{Cavendish Laboratory, Department of Physics, University of Cambridge, Cambridge CB3 0HE, United Kingdom}

\date{\today}

\begin{abstract}
    Computational chemistry is one of the most promising applications of quantum computing, mostly thanks to the development of the Variational Quantum Eigensolver (VQE) algorithm. VQE is being studied extensively and numerous optimisations of VQE's sub-processes have been suggested, including the encoding methods and the choice of excitations. Recently, adaptive methods were introduced that apply each excitation iteratively. When it comes to adaptive VQE, research is focused on the choice of excitation pool and the strategies for choosing each excitation. Here we focus on a usually overlooked component of VQE, which is the choice of the classical optimisation algorithm. We introduce the parabolic optimiser that we designed specifically for the needs of VQE. This includes both an 1-D and an n-D optimiser that can be used either for adaptive or traditional VQE implementations. We then continue to benchmark the parabolic optimiser against Nelder-Mead for various implementations of VQE. We found that the parabolic optimiser performs significantly better than traditional optimisation methods, requiring fewer CNOTs and fewer quantum experiments to achieve a given energy accuracy.
  
\end{abstract}

\maketitle

\section{\label{Introduction}Introduction}
    This paper builds on the recent research on the Variational Quantum Eigensolver (VQE) \cite{vqe_general, vqe_general_2, vqe_2,VQE_accelerated, vqe_google_hf}. VQE is a hybrid quantum-classical algorithm (HQC) algorithm, that can be used to solve the electronic structure problem \cite{vqe_general, CCSD}. The main benefit of VQE is that it requires relatively shallow quantum circuits and thus it can be successfully executed on noisy intermediate-scale quantum (NISQ) \cite{NISQ,google_supreme, vqe_nisq} computers. 
    
    Our goal during the development of this project was to optimize the implementation of VQE so that it runs as fast as possible and with the minimum amount of resources. In the following sections we describe in detail the various steps of our VQE implementation.

\section{\label{Optimiser}Parabolic Optimiser}
    As a hybrid quantum-classical algorithm, VQE consists of both a classical and a quantum part. In this section we will focus solely on the classical part. Specifically, the VQE algorithm is used to optimize a function $f(\pmb{x})$, with respect to the vector of optimization parameters $\pmb{x}$. For a given $\pmb{x}$, $f(\pmb{x})$ is calculated on a quantum computer. However, the choice of $\pmb{x}$ is left to the classical computer that does the optimisation. Here, when we refer to optimisation, we usually mean finding (or approximating) the global minimum of $f(\pmb{x})$. 

    While there are many single and multi variable optimisation techniques readily available, we did not find one that fitted all our needs. So, we went ahead and developed one of our own.
    
    \subsection{1-D Optimiser}
        Here we describe the function of our 1-D parabolic optimiser. The main application of this optimiser is for when all but one components of $\pmb{x}$ are kept constant. Therefore, $f(\pmb{x})$ can be viewed as a single parameter function $g(x_i)$. 
       
        Here, we drew inspiration from the particular nature of our problem. Specifically, each component of $\pmb{x}$ models the process of exciting one or several electrons from one state to some other one. Furthermore, this excitation does not take place in random states. Due to the nature of the algorithm, we are guaranteed that the initial state will be the Hartree Fock ground state, or potentially an even better approximation of the actual ground state of our configuration. This means that each $x_i$ is close to $0$. Also, an excitation for $x_i=0$ has no effect on the energy. 
        To recap, we are trying to optimise a variable $x_i$ and we know that the location of the minimum will be close to $0$. We also know the general form of $g(x_i)$; it is a sinusoid curve with a period of $2\pi$ and a minimum close to $0$.
       
        Finally, we know from basic physics that near a local minimum any function can be approximated as a parabola. We use this fact to approximate locally $g(x_i)$ as a parabola. The particular implementation of our 1-D optimiser follows.
       
        We define a distance ${\delta}x$ (we found that ${\delta}x=0.1$ works quite well). On the quantum computer we calculate the function $g(x_i)$ for $x_i=0+{\delta}x$ and $x_i=0-{\delta}x$. Note that we do not need to calculate $g(0)$ since, this is the energy of the previous iteration (or the Hartree Fock energy if this is the first iteration).
       
        We then identify the parabola $P(x)=a(x-x_0)^2+b$ that is defined by these three points ($0+{\delta}x$, $0-{\delta}x$, $0$). The location of the minimum can be easily shown to be:
        
        \begin{equation} \label{single_var_optimiser}
            x_0 = \frac{{\delta}x}{2} \frac{g(0+{\delta}x) - g(0-{\delta}x)}{g(0+{\delta}x) + g(0-{\delta}x) - 2g(0)}.
        \end{equation}
        
        From testing, we found that although this method works exceptionally well for identifying the location of the minimum $x_0$, it performs poorly in in the identification of the actual value of the minimum $g(x_0)$. Thus, we calculate $g(x_0)$ on the quantum computer, as we would normally. 
        
        Note, that $x_0$ being far away from $0$ implies that the parabolic approximation of $g(x)$ might not be good enough, since we are too far away from the minimum. In this case, an extra step can be added, where, $g(x)$ is approximated by a parabola near $x_0$ instead of $0$. In this case the new $x_0'$ can be calculated from \autoref{single_var_optimiser}, by replacing $0$ with $x_0$. From our testing we found that this procedure does not need to be repeated more than two times, while for the vast majority of the cases, one iteration is enough.
        
        \subsubsection{Complexity Analysis}
        
            As the reader can infer from the previous discussion, the complexity of the 1-D optimiser is $O(1)$. Specifically, for each iteration, $3$ calculations are needed. One for $g(0+{\delta}x)$, one for $g(0-{\delta}x)$ and a final one for  $g(x_0)$. Since this algorithm requires at most two iterations, in total it will need either $3$ or $6$ calculations, with $3$ being by far the most common case.
            
            With such a favorable complexity, this is the best optimisation method that we are aware of that has been implemented in the context of VQE. However, there is still a major drawback. Namely, this is a 1-D optimiser and VQE has to deal with multi-variable optimisation. We explain how we deal with this issue in the following section.
            
    \subsection{\label{n-d_optimiser}n-D Optimiser}
        
        It would be intuitive to try to generalise the above method to $n$ dimensions. In fact, one can fit an $n$-dimensional elliptical paraboloid near the minimum of $f(\pmb{x})$. The complexity of such an algorithm would be $O(n^2)$ since one would have to take into account that this is a general, rotated paraboloid. This complexity is not good enough for our needs so we had to opt for a different approach.
        
        The first thing that came to mind was to approximate the minimum with an elliptical paraboloid that has its axes along the basis vectors of $\pmb{x}$. This algorithm has a complexity of $O(n)$ since it disregards rotations. However, this algorithm is not a good enough approximation of reality, since the paraboloid is actually rotated which means that the algorithm will not necessarily converge.
        
        The solution we opted for, was the following. We calculated the gradient $\pmb{\nabla}f(\pmb{x})$. We can assume that the gradient vector points somewhere close to the minimum. We can then use the gradient to define an axis along which we can run the 1-D optimiser, to identify this minimum. 
        
        Specifically, we define:

        \begin{equation}
            h(t) = f\left(\pmb{0} - \frac{\pmb{\nabla}f(\pmb{0})}{|\pmb{\nabla}f(\pmb{0})|}t\right) .
        \end{equation}
        
        It is now sufficient to optimise the scalar function $h(t)$ in order to find the minimum of $f(\pmb{x})$
        
        Note here, that we are not guaranteed that the minimum of $f(\pmb{x})$ is along its gradient vector. However, it should be close. One then, can opt to perform a second iteration, replacing $\pmb{0}$ with $\pmb{x_0}$ to arrive to a better approximation $\pmb{x_0'}$. 
        
        In our case, for reasons that will become apparent at \autoref{total_CA}, a second iteration is not needed. 
        
        We calculate the gradient vector by finding the partial derivative with respect to each variable $x_i$ numerically using:
        
        \begin{equation}
            \frac{\partial f}{\partial x_i}\bigg|_{\pmb\kappa} = \frac{f(\pmb\kappa +\delta x \cdot \pmb{\hat{e}}_i) - f(\pmb\kappa)}{\delta x},
        \end{equation}
        
        where $\kappa$ is the point at which we want to evaluate the derivative and $\pmb{\hat{e}}_i$ the unit vector corresponding to the $i^{th}$ parameter.
        
        Here we used $\delta x=10^{-6}$ for the calculation of the derivatives. The point $\pmb\kappa$ is the result of the previous iteration (or $\pmb0$ during the first iteration). Thus, $f(\pmb\kappa)$ has already been calculated in the previous step (or is the Hartree-Fock energy during the first iteration). This means that in order to calculate the partial derivatives, we only need to calculate $f(\pmb\kappa +\delta x \cdot \pmb{\hat{e}}_i)$ for each $i$. Therefore, exactly one expectation value measurement is needed in order to calculate each partial derivative.
        
        \subsubsection{Complexity Analysis}
            
            The most computationally intensive task required so far is the calculation of the gradient vector. Since this is an $n$-dimensional vector, $n$ calculations are needed on the quantum computer, thus the computational complexity is $O(n)$. Note however that all these calculations are independent from each other, allowing for all these tasks to be performed in parallel. 
            
            Taking all into account, for a system with $n$ excitations this algorithm requires between $n+3$ and $2n+6$ calculations to be performed on a quantum computer, depending on the user's requirements. In our case, as will be shown in \autoref{total_CA}, we require $n+3$ iterations most of the time, and $n+6$ in some extreme cases.
         
\section{\label{Adapt_VQE}Adaptive VQE}
    
    A significant advancement in the field of VQE is the introduction of adaptive algorithms \cite{ADAPT, Q-ADAPT, gene_vqe, iterative_QCC, iQCC_ILC, pruning_vqe, benchmark_adapt_vqe, iter_vqe_6} that build the quantum circuit step-by-step. All these algorithms introduce a pool of potential excitations and employ a strategy to choose the next excitation based on the results achieved in the previous step. This procedure is repeated until the energy converges. Note here that once an excitation has been picked, it is not removed from the pool. Instead, it can be chosen again, should the algorithm decide to. 
    
    Throughout this process more and more optimisation parameters are introduced, (usually one per excitation). It is therefore clear that the choice of optimisation algorithm will impact the performance of the whole process. In our case, we opted to use our n-D parabolic optimiser, described in \autoref{n-d_optimiser} but other popular choices include the Nelder-Mead and BFGS optimisation algorithms.
    
    \subsection{\label{selection}Strategies for Selecting Excitations}
    
        In the original ADAPT-VQE paper \cite{ADAPT}, the authors use as a metric for the best excitation the gradient of the energy with regards to each excitation parameter. More specifically, they choose to perform the excitation with the largest gradient. Although this is a very intuitive approach, we found that it has several significant drawbacks. Specifically, the gradient can be defined as: 
        
        \begin{equation} 
                \frac{\partial E}{\partial x_i} = \langle\Psi|[\hat{H},\hat{A_i}]|\Psi\rangle,
        \end{equation}
        
        where, $\hat{A_i}$ is the $i$th excitation operator and $\hat{H}$ the Hamiltonian. In order to calculate this on a quantum computer, one has to calculate the operator $[\hat{H},\hat{A_i}]$. Because both $\hat{H}$ and $\hat{A}$ are written in the form:
        
        \begin{equation} \label{ham_second_quant}
            \hat{H} = \sum_{pq}h_{pq}\alpha_{p}^\dagger\alpha_{q}+
                    \frac{1}{2}\sum_{pqrs}h_{pqrs}\alpha_{p}^\dagger\alpha_{q}^\dagger\alpha_{r}\alpha_{s},
        \end{equation}
        
        finding analytically the commutator on a classical computer has a high computational cost since it requires matrix multiplications. However, there is a less expensive way of calculating the derivatives. Specifically, one can use the approximation:
        
        \begin{equation} 
            \frac{df}{dx} \approx \frac{f(x+\delta x) - f(x)}{\delta x}.
        \end{equation}
        
        Since $f(x)$ is known, only $f(x+\delta x)$ is needed. Thus the complexity for calculating all derivatives becomes $O(n)$, requiring exactly $n$ operations on a quantum computer. This can also be parallelised and can be run on multiple quantum computers at the same time. 
        
        However, our testing showed that using gradients is not the optimal approach for the choice algorithm, since the excitation with the largest gradient is not guaranteed to have the greatest impact on the energy. 
        
        Instead, what we opted to do, is to identify the excitation that reduces the energy by the most significant margin. This can be done by using our 1-D optimiser, to calculate the impact of each excitation. 
        
        Specifically, say we have reached a state $|\Psi\rangle$. For each excitation $\hat{A_i}$ in our pool we evaluate the expectation value
        
        \begin{equation} \label{exp_value}
            E_i = \langle\Psi'(x_i)|\hat{H}|\Psi'(x_i)\rangle
        \end{equation}
        
        on a quantum computer. Here, $|\Psi'(x_i)\rangle = \hat{A_i}(x_i)|\Psi\rangle$. 
        
        We then run our 1-D optimiser for each one of the $\hat{A_i}$s in our pool and keep the one that produces the smallest energy $E_i$. This becomes our choice of excitation.
        
        Note here, that although both the derivative based and the energy based approaches are of complexity $O(n)$, the energy approach requires $\sim3n$ iterations, while the derivative approach requires $n$. However, because the energy approach converges faster, it will require fewer iterations for a given accuracy and therefore it will make the global optimisation step more efficient. Also, since it requires fewer excitations in total, it will produce a circuit with fewer CNOTs, which is of paramount importance.
        
        As a final note, in this paper we consider the expectation value measurements of \autoref{exp_value} to be the fundamental operation performed on a quantum computer. Therefore, when we use the term "quantum experiment" we actually refer to an expectation value measurement. These two terms are used here interchangeably since we do not delve into the particular procedures required to extract the expectation value of the Hamiltonian. In terms of the implementation of the algorithm, this is achieved using Qiskit's statevector simulator \cite{Qiskit} and the matrix representation of the Hamiltonian. Thus \autoref{exp_value} becomes $E = \pmb{u}^\dagger \hat{H} \pmb{u}$, where $\pmb{u}$ is the corresponding statevector.
    
    \subsection{\label{total_CA}Complexity Analysis}
    
        Once we have picked the right excitation for the next step, it is time to perform a global optimisation between all the excitations that have been selected so far. This is done with our $n$-dimensional parabolic optimiser.
        
        Keep in mind that each time an excitation is chosen, the dimensionality of the search space is increased by one. Thus, for each iteration, the search space is a subspace of the search space of the next iteration. This allows us to run the $n$-dimensional parabolic optimiser only once per excitation, although we are not guaranteed that we will find the local minimum. This is sufficient because once an additional excitation is added, the optimiser will search again the same (albeit expanded) search space.
        
        If we consider an algorithm with $M$ potential excitations, the $n^{th}$ iteration will require $\sim n+3M$ distinct executions on the quantum computer. Therefore, if the algorithm converges after $n$ iterations, it will have required in total $\sim\frac{n(n+1)}{2} +3nM$ quantum experiments. Thus we have arrived to a very efficient algorithm with a complexity of $O(n^2)$. 
        
        As a final note for this section, the maximum CNOT count of this algorithm is $\sim nk$, where $k$ is the average CNOT count between the $M$ excitations in our pool. It has been shown in \cite{ADAPT} that for a given accuracy ADAPT-VQE vastly reduces the CNOT count compared to a typical VQE implementation.
        
\section{Circuit Depth Reduction}

    So far we have focused our attention mainly on the classical part of the VQE algorithm. Here, we will shift our focus towards improving its quantum part. Note that from here on out, the classical complexity will remain roughly the same. Instead, we will focus on reducing the CNOT count.
    
    Here, we need to specify the encoding that we are going to use for representing the molecular Hamiltonian at a quantum hardware level. Popular choices include the Jordan-Wigner \cite{JW_encode} or the Bravyi-Kitaev \cite{BK_encode} encoding. We opted to use the Jordan-Wigner encoding since it provides a useful one-to-one mapping between orbitals and qubits. As for the excitations, we are going to use the popular unitary coupled cluster (UCC) \cite{vqe_first_uccsd} ansatz, keeping only single-electron and double-electron excitations (UCCSD) \cite{UCCSD, UCCSD_2, UCCSD_3, UCCSD_4, UCCSD_bogoliubov, optimized_uccsd}.
    
    Under the Jordan-Wigner encoding the the UCC single excitations take the form
    
    \begin{equation}\label{eq:single_exc}
        \exp\big[i\frac{\theta}{2}(X_i Y_k - Y_i X_k)\prod_{r=i+1}^{k-1}Z_r \big].
    \end{equation}
    
    Similarly, the double excitations can be written as
    
    \begin{multline}
        \label{eq:double_exc}
        \exp \big[ i\frac{\theta}{8} (X_i Y_j X_k X_l + Y_i X_j X_k X_l + Y_i Y_j Y_k X_l \\  +  Y_i Y_j X_k Y_l  - X_i X_j Y_k X_l  - X_i X_j X_k Y_l \\ - Y_i X_j Y_k Y_l - X_i Y_j Y_k Y_l ) \prod_{r=i+1}^{j-1} Z_r\prod_{r'=k+1}^{l-1} Z_{r'}\big].
    \end{multline}
    
    \subsection{\label{spin}Spin Conservation}
    
        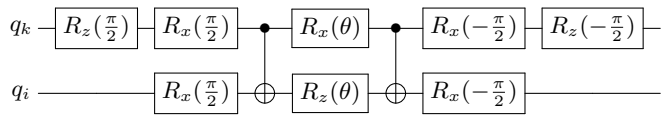
\begin{figure}[t]
            \Qcircuit @C=0.7em @R=1.0em {
            q_k && \gate{R_z(\frac{\pi}{2})} & \gate{R_x(\frac{\pi}{2})} & \ctrl{1} & \gate{R_x(\theta)} & \ctrl{1}  & \gate{R_x(-\frac{\pi}{2})} & \gate{R_z(-\frac{\pi}{2})} & \qw \\
            q_i &&  \qw & \gate{R_x(\frac{\pi}{2})} & \targ & \gate{R_z(\theta)} & \targ & \gate{R_x(-\frac{\pi}{2})} & \qw & \qw \\
            }
            \centering
            \caption{Circuit  implementing the single qubit excitation of matrix \ref{2qubit_exc}}
            \label{fig:2qubit_circ}
        \end{figure} 
        
        \begin{figure*}[p]
            \centering
            \includegraphics[width=0.95\textwidth]{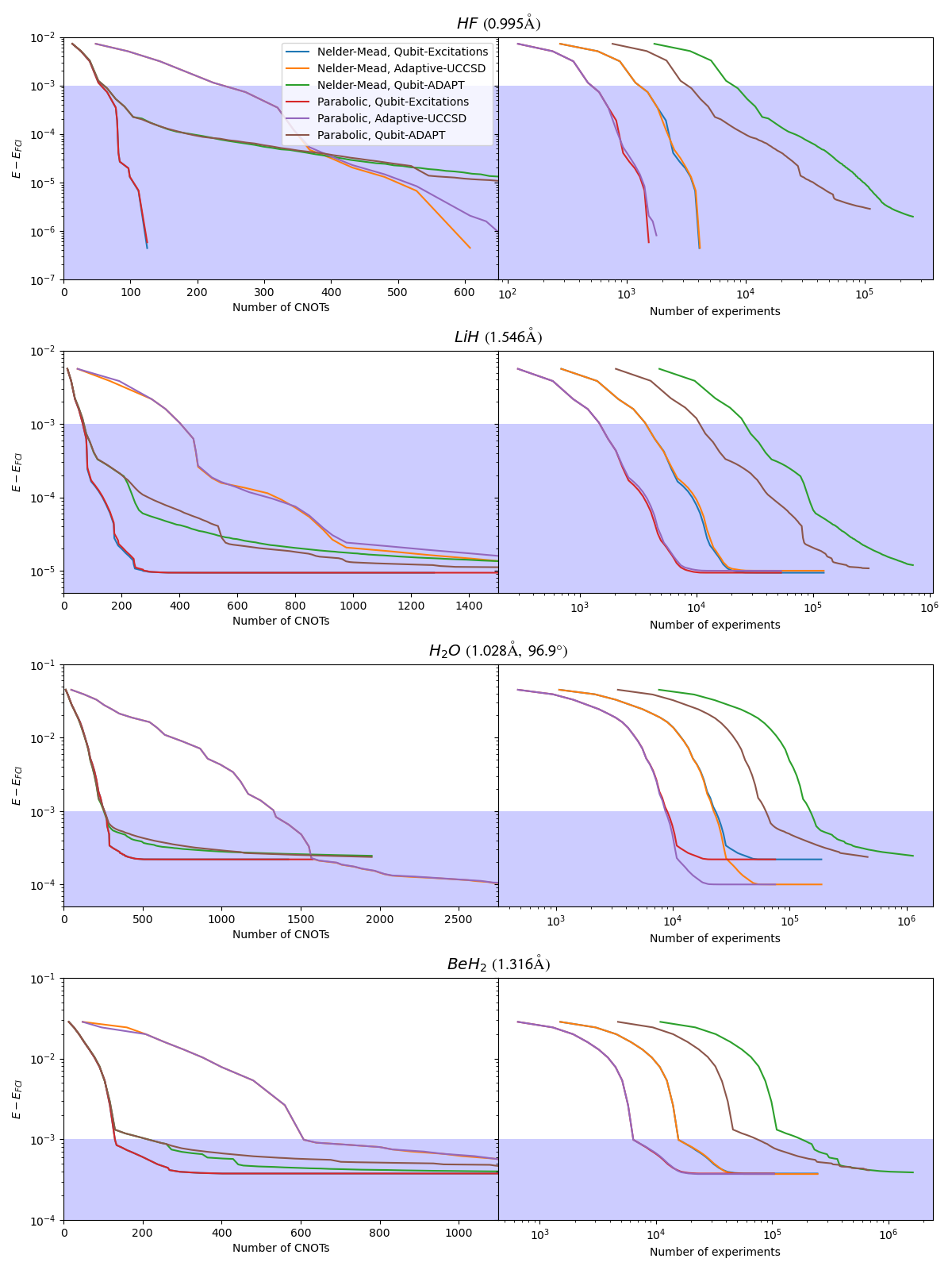}
        
            \captionsetup{width=1.0\textwidth,justification=centerlast,singlelinecheck=on}
        
            \caption{\textbf{Simulations of several different molecules using different variations of VQE.} For all molecules that were simulated, the parabolic Qubit-Excitations and Nelder-Mead Qubit-Excitations yielded the least number of CNOTs, with nearly identical results. At the same time, the least number of experiments was produced by parabolic Qubit-Excitations and parabolic Adaptive-UCCSD, again with quite similar results. Since our objective is to minimize both the number of CNOTs and the number of experiments simultaneously, the parabolic Qubit-Excitations method is the strongest contender.}
            \vspace{-0.2cm}
            \rule{\linewidth}{0.5pt}
            \label{fig:results}
        \end{figure*}
         
        \begin{figure*}[t]
            \Qcircuit @C=0.3em @R=.75em {
            &&& \\ 
            q_l \ && \ctrl{1} & \qw & \ctrl{2} & \gate{R_y(\frac{\theta}{8})} & \ctrl{1} & \gate{R_y(-\frac{\theta}{8})} & \ctrl{3} & \gate{R_y(\frac{\theta}{8})} & \ctrl{1} & \gate{R_y(-\frac{\theta}{8})} & \ctrl{2} & \gate{R_y(\frac{\theta}{8})} & \ctrl{1} & \gate{R_y(-\frac{\theta}{8})} & \ctrl{3} & \gate{R_y(\frac{\theta}{8})} & \ctrl{1} & \gate{R_y(-\frac{\theta}{8})} & \ctrl{2} &  \gate{R_z(\frac{\pi}{2})} & \qw & \ctrl{1} & \qw \\
            q_k \ && \targ{} & \gate{X} & \qw & \gate{H} & \targ & \qw & \qw & \qw & \targ & \qw & \qw & \qw & \targ & \qw & \qw & \qw & \targ & \gate{H} & \qw & \qw & \gate{X} & \targ & \qw & \\
            q_j \ && \ctrl{1} & \qw & \targ{} & \qw & \qw  & \qw & \qw & \qw & \qw & \gate{H} & \targ & \qw & \qw  & \qw & \qw & \qw & \qw & \gate{R_z(-\frac{\pi}{2})} & \targ & \gate{R_z(-\frac{\pi}{2})}& \gate{R_y(-\frac{\pi}{2})}  & \ctrl{1} & \qw & \\
            q_i \ && \targ{} & \gate{X} & \qw & \qw & \qw & \gate{H} & \targ & \qw & \qw & \qw & \qw & \qw & \qw & \qw & \targ & \gate{H} & \qw & \qw & \qw  & \qw &\gate{X} & \targ & \qw &  \\
            }
            \caption{Circuit implementing the double qubit excitation  of matrix \ref{4qubit_exc}}
            \label{fig:4qubit_circ}
        \end{figure*}
    
        One of the easiest optimisations one can perform is to reduce the size of the excitation pool. Specifically, we opted to disregard any and all excitations that do not conserve spin. This significantly reduces the number of excitation parameters, with no impact on the energy accuracy for the specific molecular configurations that we chose.
        
        In the Jordan-Wigner encoding spin-down orbitals are mapped to even-numbered states and spin-up orbitals to odd-numbered states. This means that for the case of the single excitations of \autoref{eq:single_exc} we only keep the excitations that satisfy $i \bmod 2 = k \bmod 2$, where $\bmod$ denotes the modulo operation. Similarly, for the double excitations of \autoref{eq:double_exc} we require that $(i \bmod 2) + (j \bmod 2) = (k \bmod 2) + (l \bmod 2)$.
    
    \subsection{\label{Qubit-ADAPT pool}The Qubit-ADAPT pool}
    
        One approach to reduce the circuit depth further, is that of Qubit-ADAPT \cite{Q-ADAPT}. This approach introduces Pauli string exponentials that practically correspond to each term of \autoref{eq:single_exc} and \autoref{eq:double_exc} after dropping the $Z$ operators. More specifically, single  and double excitations are given by
        
        \begin{equation}
            \exp \big[ i \theta \sigma_i \sigma_k \big ]
        \end{equation}
        
        and
        
        \begin{equation}
            \exp \big[ i \theta \sigma_i \sigma_j \sigma_k \sigma_l \big ] ,
        \end{equation}
        
        respectively. Here $\sigma$ is to be replaced by either $X$ or $Y$ and each excitation should contain an odd number of $Y$s to satisfy orthogonality (unitarity). 
        
    \subsection{\label{Qubit-Excitation pool}The Qubit-Excitation pool}

        Our approach was to just omit the Pauli $Z$ strings from the UCCSD excitations, thus introducing what we call Qubit-Excitations \cite{old, new}. This means that \autoref{eq:single_exc} becomes:
        
        \begin{equation}\label{eq:single_exc_no_Z}
            \exp\big[i\frac{\theta}{2}(X_i Y_k - Y_i X_k)\big].
        \end{equation}
        
        Similarly, \autoref{eq:double_exc} becomes
        
        \begin{multline}
            \label{eq:double_exc_no_Z}
            \exp \big[ i\frac{\theta}{8} (X_i Y_j X_k X_l + Y_i X_j X_k X_l + Y_i Y_j Y_k X_l +  Y_i Y_j X_k Y_l \\ - X_i X_j Y_k X_l  - X_i X_j X_k Y_l - Y_i X_j Y_k Y_l - X_i Y_j Y_k Y_l ) \big].
        \end{multline}
        
        The main benefit of this pool is that all single and double excitations correspond to the same operator, just acting on different qubits. The same is true for triple excitations quadruple excitations and so on. 
        
        In the following, we will focus solely on single and double excitations although we would be excited (no pun intended) to see progress for higher degree excitations. 
        
        The excitations of \autoref{eq:single_exc_no_Z} and \autoref{eq:double_exc_no_Z} can also be written in their matrix form, that in the computational basis corresponds to the matrices \ref{2qubit_exc} and \ref{4qubit_exc} respectively. Note that these are orthogonal matrices that belong to $SO(4)$ and $SO(16)$ respectively.
         
        \begin{equation} \label{2qubit_exc}
        \begin{pmatrix}
            1 &   0    &    0    &  0 \\
            0 & cos(x) & -sin(x) &  0 \\
            0 & sin(x) &  cos(x) &  0 \\
            0 &   0    &    0    &  1
        \end{pmatrix}
        \end{equation} 
        
        \begin{multline} \label{4qubit_exc}
        \begin{pmatrix}
            1 & 0 & 0 & 0 & 0 & 0 & 0 & 0 & 0 & 0 & 0 & 0 & 0 & 0 & 0 & 0 \\
            0 & 1 & 0 & 0 & 0 & 0 & 0 & 0 & 0 & 0 & 0 & 0 & 0 & 0 & 0 & 0 \\
            0 & 0 & 1 & 0 & 0 & 0 & 0 & 0 & 0 & 0 & 0 & 0 & 0 & 0 & 0 & 0 \\
            0 & 0 & 0 & cos(x) & 0 & 0 & 0 & 0 & 0 & 0 & 0 & 0 & -sin(x) & 0 & 0 & 0 \\
            0 & 0 & 0 & 0 & 1 & 0 & 0 & 0 & 0 & 0 & 0 & 0 & 0 & 0 & 0 & 0 \\
            0 & 0 & 0 & 0 & 0 & 1 & 0 & 0 & 0 & 0 & 0 & 0 & 0 & 0 & 0 & 0 \\
            0 & 0 & 0 & 0 & 0 & 0 & 1 & 0 & 0 & 0 & 0 & 0 & 0 & 0 & 0 & 0 \\
            0 & 0 & 0 & 0 & 0 & 0 & 0 & 1 & 0 & 0 & 0 & 0 & 0 & 0 & 0 & 0 \\
            0 & 0 & 0 & 0 & 0 & 0 & 0 & 0 & 1 & 0 & 0 & 0 & 0 & 0 & 0 & 0 \\
            0 & 0 & 0 & 0 & 0 & 0 & 0 & 0 & 0 & 1 & 0 & 0 & 0 & 0 & 0 & 0 \\
            0 & 0 & 0 & 0 & 0 & 0 & 0 & 0 & 0 & 0 & 1 & 0 & 0 & 0 & 0 & 0 \\
            0 & 0 & 0 & 0 & 0 & 0 & 0 & 0 & 0 & 0 & 0 & 1 & 0 & 0 & 0 & 0 \\
            0 & 0 & 0 & sin(x) & 0 & 0 & 0 & 0 & 0 & 0 & 0 & 0 & cos(x) & 0 & 0 & 0 \\
            0 & 0 & 0 & 0 & 0 & 0 & 0 & 0 & 0 & 0 & 0 & 0 & 0 & 1 & 0 & 0 \\
            0 & 0 & 0 & 0 & 0 & 0 & 0 & 0 & 0 & 0 & 0 & 0 & 0 & 0 & 1 & 0 \\
            0 & 0 & 0 & 0 & 0 & 0 & 0 & 0 & 0 & 0 & 0 & 0 & 0 & 0 & 0 & 1
        \end{pmatrix}
        \end{multline} 
    
    \subsection{\label{Circuit Rewriting}Circuit Rewriting}
    
        Since we have exactly two unique matrices that we will ever have to implement as circuits, we can now afford to work more towards optimising their implementation. Contrast this to the circuits arising from Qubit-ADAPT that include a total of ten unique circuits. Also, note that this would be practically impossible in the case of the circuits of \autoref{eq:single_exc} and \autoref{eq:double_exc}, since the $Z$-terms that arise mean that for each excitation in the pool we would have to optimise a unique circuit.
    
        In our case, we have shown in \cite{old} that the matrices \ref{2qubit_exc} and \ref{4qubit_exc} can be represented by the circuits of \autoref{fig:2qubit_circ} and \autoref{fig:4qubit_circ} respectively. Thus the simulations that follow require 2 CNOTs for single Qubit-Excitations and 13 CNOTs for double Qubit-Excitations. At the same time, the single and double Qubit-ADAPT excitations require 2 and 6 CNOTs respectively. Although at first glance it appears like Qubit-ADAPT requires a smaller number of CNOTs compared to Qubit-Excitations, that is not true. That is because Qubit-ADAPT will converge more slowly, thus for a given energy accuracy the overall CNOT count tends to be much higher (see also \autoref{fig:results}).
        
\section{Results}
    In \autoref{fig:results} we present the results we got from simulating $HF$, $LiH$, $H_2O$ and $BeH_2$. In all simulations we used the STO-3G basis and we only accounted for excitations that preserve spin. For each run, we used both our own n-D parabolic optimizer as well as the Nelder-Mead optimizer which is a common choice for VQE implementations. For both choices of optimizer we run VQE for three different choices of excitation pool. We used the Qubit-ADAPT pool as outlined in \autoref{Qubit-ADAPT pool}, the Qubit-Excitation pool of \autoref{Qubit-Excitation pool} and a third pool that we refer to as Adaptive-UCCSD. The Adaptive-UCCSD pool consists of the UCC single and double excitations given by \autoref{eq:single_exc} and \autoref{eq:double_exc} respectively. Therefore, Adaptive-UCCSD pool consists of the usual UCCSD excitations but these excitations are performed in an adaptive manner. This serves as an additional benchmark of our optimiser, that demonstrates its performance under a more common VQE implementation. Note here that Adaptive-UCCSD excitations are implemented through the usual staircase operators. For all three pools, we have not taken into account potential cancelling out of neighbouring CNOTs nor used other potential circuit optimisation techniques.
    
    \autoref{fig:results} helps us draw several conclusions about the various methods that were used. The parabolic optimiser outperforms Nelder-Mead in terms of number of experiments that are required to reach a certain energy accuracy. This happens while there is no significant impact on the number of CNOTs required to reach this accuracy. At the same time, we can also conclude that the choice of pool significantly affects the number of CNOTs required, with the smallest number of qubits achieved when using the Qubit-Excitation pool. Moreover, it is worth mentioning that the Qubit-ADAPT pool produces a significantly higher number of experiments compared to the Qubit-Excitation and Adaptive-UCCSD pools. That is because the Qubit-ADAPT pool consists of many more excitations than the other pools (one UCC double excitation corresponds to eight Qubit-ADAPT excitations).

\section{Outlook and Discussion}
    The parabolic Qubit-Excitation algorithm appears to be the most promissign of the VQE implementations that we tested. It has a complexity of $O(n^2)$ as it requires $\sim\frac{n(n+1)}{2} +3nM$ distinct experiments to be performed on a quantum computer. Most of these experiments can be run in parallel. As for the CNOT count, it requires $\sim 13n$ CNOTs where n is the number of iterations.
        
    It would be very useful to try to reduce the circuit depth even further. This could be done by aiming to reduce the number of iterations required to get to the desired accuracy. Adding higher degree excitations, like triple excitations is one way to achieve that but the corresponding circuits would then have to be optimised as well. A comprehensive theory for the minimal CNOT decomposition of such orthogonal matrices would go a long way.

\bibliography{bib}
\end{document}